# High-contrast two-quantum optically detected resonances in NV centers in diamond in zero magnetic field

A K Dmitriev and A K Vershovskii

*Ioffe Institute, Russian Academy of Sciences, St. Petersburg, 194021 Russia*
*e-mail address: antver@mail.ioffe.ru*

The methods for controlling spin states of negatively charged nitrogen-vacancy centers using a combination of microwave (MW) or radiofrequency (RF) excitation field for electron spin transitions and RF excitation field for nuclear spin transitions are most effective in strong magnetic fields where level anti-crossing (LAC) occurs. However, LAC in zero field can also be used to control spin states, as well as to excite narrow resonances for metrological application. In this paper we present magnetically independent resonances arising in the ODMR spectra of NV centers in bulk diamond under two-frequency (MW+RF) resonant excitation in zero magnetic field, and discuss their specificity.

**Keywords:** Nitrogen-vacancy centers, optical detection, zero-field level anti-crossing, two-quantum resonance

## INTRODUCTION

The optically detected magnetic resonance (ODMR) in negatively charged nitrogen-vacancy (NV) center in diamond has been thoroughly studied over the past decades. Many types of optically detected resonances, both magnetically dependent and magnetically independent, were discovered and investigated, and various methods of controlling both electron and nuclear spins were developed. These investigations have resulted in the development of new methods of high-resolution quantum magnetometer [1-3], as well as methods of frequency stabilization [4]. The techniques of controlling NV center electron and nuclear spins are considered to be highly suitable for quantum information processing, because the relaxation time of these spins is very long compared to other solid-state objects [5,6]. Typically, either combined microwave (MW) and radiofrequency (RF) excitation [7-14], or electron spin-echo envelope modulation methods [15] are used to address a chosen spin state, allowing for various multi-quantum resonances most thoroughly studied in [12]. Various schemes of hole-burning using multi-frequency MW and RF excitation were discussed in [14,16]; methods of all-optical excitation ODMR at two frequencies separated by the ground-state zero-field splitting were proposed in [17].

The multi-frequency methods are most effective in strong magnetic fields (0.05 or 0.1 T), where level anti-crossing (LAC) [5,6,12,16] in excited or ground state occurs. However, LAC also occurs in zero magnetic field [18], and possibilities of using it for controlling electron and nuclear spin states, or for exciting narrow resonances for metrological applications are not yet sufficiently explored.

Here we present the results of investigating ODMR spectra of NV centers in bulk diamond in zero magnetic fields using two-frequency (MW+RF) resonance excitation, and demonstrate high-contrast two-quantum resonances due to the zero-field LAC. The contrast and steepness of these resonances exceed these of ordinary ODMR, and their frequency is independent of the magnetic field's value, being entirely defined by axial zero-field splitting. This makes them a perspective object for frequency and time metrology.

## ENERGY STRUCTURE OF THE GROUND-STATE OF NV CENTER

The level structure of $^3A_2$ ground state in external magnetic field $\vec{B}$ is defined by the Hamiltonian [19,20]:

$$H = D(S_z^2 - \frac{1}{3}\vec{S}^2) + E(S_x^2 - S_y^2) + g_s\mu_B\vec{B}\cdot\vec{S} + \\ + A_{\parallel}S_zI_z + A_{\perp}(S_xI_x + S_yI_y) + PI_z^2 - g_I\mu_N\vec{B}\cdot\vec{I}, \quad (1)$$

where $\mu_B = h\cdot13.996\cdot10^9$ Hz/T is the Bohr magneton, $\vec{I}$ is the $^{14}$N nuclear spin ($I = 1$), $\vec{S}$ is the electron spin of NV center ($S = 1$), $\mu_N = h\cdot7.622\cdot10^6$ Hz/T is the nuclear magneton, $D = 2.87$ GHz and $E$ are axial and transverse zero-field splitting (ZFS) parameters, $g_s = 2.003$ and $g_I = 0.403$ are electron and nuclear g-factors, $A_{\parallel} = -2.16$ MHz and $A_{\perp} = -2.7$ MHz are axial and transverse hyperfine splitting parameters, $P = -4.95$ MHz is the quadrupole splitting parameter. Denote eigenstates of the ground state $|m_S, m_I\rangle$; for the nitrogen isotope $^{14}$N both electronic and nuclear spin projections take values $m_S$, $m_I = 0, \pm1$.

The energy structure of NV center in zero and ultra-weak fields is more complex than in strong ones (Fig.1); it contains both level crossings and anti-crossings, partially masked by the inhomogeneity of the crystal's internal fields. Each NV center in a bulk sample is affected by local magnetic and electric fields. At $B \approx 0$, the transverse component of the electric field combined with the strain causes $m_S = \pm1$ states to mix in superpositions, and transverse magnetic field causes second-order LAC [18].

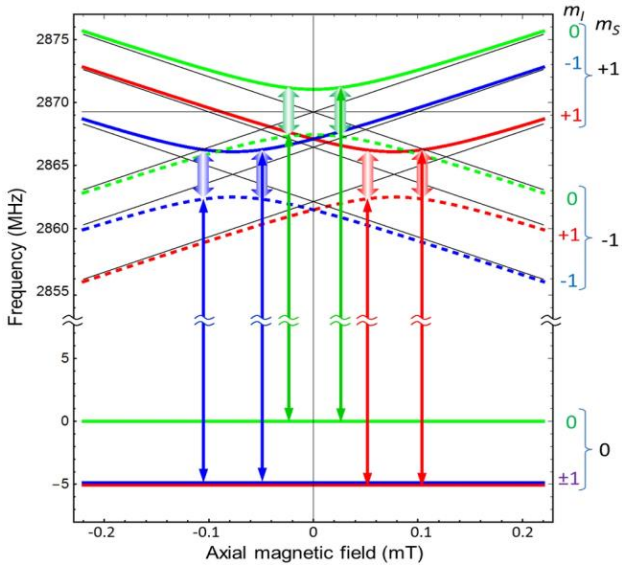

FIG. 1. (Color online) NV center ground-state splitting frequencies' dependence on axial local magnetic field, calculated for a diamond crystal with transverse ZFS parameter $E = 1.8$ MHz. Thin colored lines represent pure quantum states, thick lines are mixed states. Single arrows represent the MW drive field, double arrows represent the RF drive field.

The dependence of energy levels and corresponding frequencies on the magnetic field $B$ in low fields is substantially nonlinear. This structure was investigated in detail in [18].

## EXPERIMENTAL

The experimental setup was described in [21]: a synthetic diamond of SDB1085 60/70 grade (manufactured by Element Six) with dimensions $0.1 \times 0.3 \times 0.3$ mm was subjected to electron irradiation ($5 \cdot 10^{18}$ cm$^{-2}$) and subsequent annealing in Ar at 800°C for 2 hours. The crystal was used at a room temperature; it was fixed with an optically transparent glue to the end of an optical fiber with a core diameter of 0.9 mm; the wide fiber aperture ensured effective collection of the PL signal at the cost of losing ~ 90% of pumping light. The pumping and detection efficiencies were further increased by covering the outer surfaces of the diamond and the end of the optical fiber with a non-conductive reflective coating (Fig.2a).

The pumping beam (~15 mW at a wavelength of 532 nm) was focused on the second end of the fiber, and the PL signal was collected from the same end. Given an overall pumping efficiency of less than 10%, the pumping power was far below the optimum [10]. We excited two-frequency ODMR in $B = (0-1)$ mT using MW drive field $f_{MW}$ in combination with additional RF field $f_{RF}$. We used low-frequency amplitude modulation of RF field in order to subtract the fluorescence background.

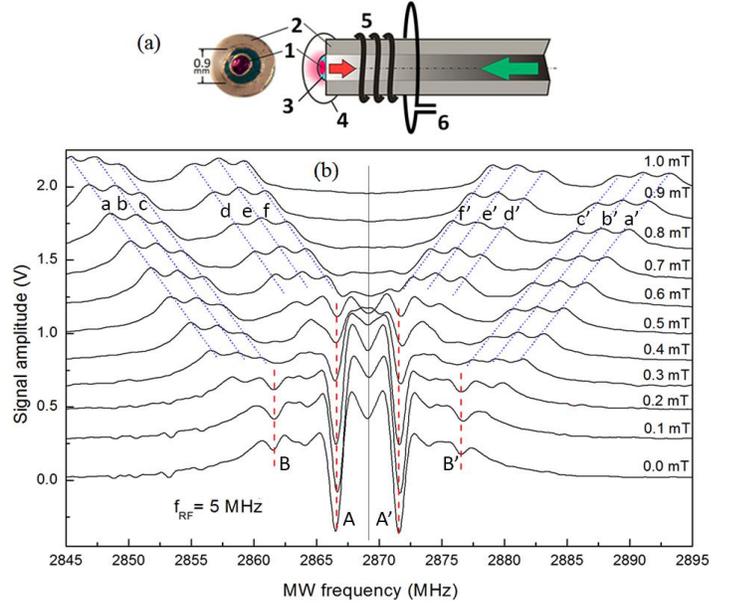

FIG. 2. (Color online) (a) View of the fiber end, and a schematic diagram of magnetometer sensor: 1 – diamond crystal, 2 – optical fiber, 3 – transparent glue, 4 – reflective coating, 5 – MW antenna, 6 – RF antenna. (b) ODMR spectra recorded at external field $B = (0-1)$ mT along $(1,0,0)$ direction with additional RF excitation at $f_{RF} = 5$ MHz; dashed and dotted lines denote multi-quantum ODMR resonances; the symmetrical hollows $AA'$ are the two-quantum resonances in question.

Under this setup, we recorded two-quantum resonances $a$-$f$, $f'$-$a'$ (Fig.2b), arising under condition $f_{MW} \pm f_{RF} = f_{ODMRi}$ ($f_{ODMRi}$ are the frequencies of one-quantum ODMR). We also recorded symmetrical hollows $AA'$, $BB'$ (Fig.2b,3), arising in ODMR spectra under conditions

$$f_{MW} \pm \tfrac{1}{2} n f_{RF} = D, \qquad (2)$$

$$\begin{aligned} v_0 - \Delta &< 2|f_{MW} - D| < v_0 + \Delta, \text{ or} \\ v_0 - \Delta &< f_{RF} < v_0 + \Delta, \end{aligned} \qquad (3)$$

where $n = 1$ for $AA'$ and $n = 3$ for $BB'$ resonances, $v_0 = (4.34 \pm 0.02)$ MHz is the center of the resonance envelope in RF scale, and $\Delta = (2.14 \pm 0.04)$ MHz is the half-width of the envelope. We have observed similar resonances previously, while applying low-frequency amplitude modulation to the MW field in order to subtract the fluorescence background [22]; this time we applied the same modulation to the RF field, which unexpectedly caused at least a two-fold increase in the contrast of the resonances, and changed their shape (inset on Fig.3b).

The remainder of this article discusses the $AA'$ resonances only. As illustrated by Fig 3, their amplitude is of the same order of magnitude as that of joint ODMR spectra in zero field (which is, in turn, several times larger than that of isolated ODMR lines).

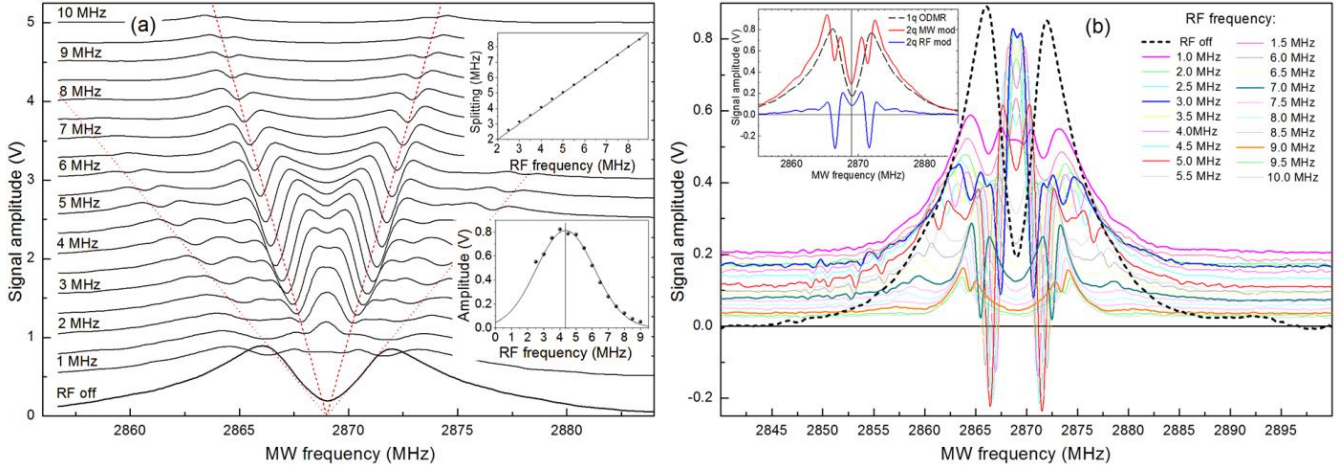

FIG. 3. (Color online)  (a) ODMR spectra recorded at zero external field at different radiofrequencies $f_{RF}$; lowest curve is normal ODMR signal, two symmetrical hollows are two-quantum resonances arising when conditions (2), (3) are fulfilled. Inset: Dependence of frequency splitting and resonance amplitudes on RF frequency. (b) close view of the same spectra. Inset: two-quantum (2q) resonances at RF and MW amplitude modulation as compared to the normal (1q) ODMR spectra.

The linewidth of $AA'$ resonances is several times smaller than that of joint ODMR spectra: at the optimal (i.e. providing the maximal resonance steepness) values of MW and RF amplitudes, the linewidth (HWHM) of the resonances was found to be about 1.1 MHz; when both amplitudes tend to zero, the resonances' width tends to $\Gamma_0/2\pi = (0.27 \pm 0.07)$ MHz. The main difference between these resonances and the resonances described in [13] is that their splitting is proportional to the frequency of the RF field, not to its power.

Also noteworthy is the nontrivial dependence of the type of ODMR signal on the modulation method. Two modulation options (RF and MW 100% amplitude modulation) are shown on the inset (Fig.3b). We also tried other combinations and found that if modulation is applied to RF and MW fields simultaneously, the signal shape is the same as in the case of MW modulation; if modulation is applied to RF and MW fields in opposite phases, the signal looks like a normal (i.e. obtained under single-frequency MW excitation) ODMR signal without any dips.

The amplitude of the peaks is maximal at zero magnetic field, and it decreases quickly as the field induction increases (Fig.2); the resonances vanish at $|B| \approx 0.5$ mT. The frequencies of both resonances proved to be insensitive to $B$; this makes them inappropriate for magnetometrical application. On the other hand, the fact that according to (2) the combination of resonant MW and RF frequencies depends only on $D$ makes these peaks very attractive for the task of frequency stabilization. The use of NV centers for timekeeping was suggested in [4]; here we propose two-quantum resonances for this purpose.

Let us estimate the sensitivity of our scheme. The peaks' amplitude at the optimum reaches 1.4 V, which corresponds to a change in the photocurrent $\Delta I_{Iph} = 1.2$ mkA, and to a relative change in the fluorescence signal at the center of the resonances $\Delta I_{Iph}/I_{Iph} = 0.027$. The shot noise level corresponding to $I_{ph0}$ is $I_{shot} = \sqrt{(2 \cdot e \cdot I_{ph} \cdot \Delta f)} = 3.75$ pA·Hz$^{1/2}$.

Therefore, the shot-noise-limited signal-to-noise ratio achievable in this configuration in 1 Hz bandwidth is about $\Delta I_{Iph}/I_{shot} = 3.2 \cdot 10^5$, or 110 dB, and the resonance linewidth (HWHM) is 1.1 MHz. It follows that by suppressing laser noise down to the shot noise level in the scheme described above, one can obtain [4] sensitivity to MW frequency variations equal to $\delta f = 3.44$ Hz$^{1/2}$, or $\delta f/f = 1.2 \cdot 10^{-9}$ Hz$^{-1/2}$. By using both peaks, the fractional instability about $0.85 \cdot 10^{-9}$ Hz$^{-1/2}$ can be achieved; this is close to what a good quartz can provide. The parametric dependencies and limitations of short-term sensitivity and long-term stability of schemes based on ODMR resonances were thoroughly investigated in [1, 23]; they are mostly applicable to the scheme proposed here. Expected long-term stability is largely limited by the temperature shift of $D$, equal to $-72.4$ kHz/K [23], and by pressure shift. Nonetheless, due to the presumably rather high short-term sensitivity achievable in a small volume, these resonances can be widely used in timekeeping schemes.

## DISCUSSION

The physics of these resonances is yet to be explained. Our calculations show that, for a given diamond sample, the frequency value $v_0$ corresponds to the ground-state splitting due to LAC, which in zero magnetic field mixes the states $|-1, m_I\rangle$ with $|+1, m_I\rangle$, where $m_I = -1, 0, 1$. Now we can rewrite (2), (3) as follows:

$$f_{MW} \pm f_{RF} = f_{(|0, m_I\rangle \leftrightarrow |\pm 1, m_I\rangle)}, \quad (4)$$

$$f_{MW} \approx f_{(|0, m_I\rangle \leftrightarrow |\mp 1, m_I\rangle)}, \quad (5)$$
$$f_{RF} \approx f_{(|\pm 1, m_I\rangle \leftrightarrow |\mp 1, m_I\rangle)}.$$

From the above, it becomes apparent that (4), (5) are conditions for a forbidden two-quantum resonance between $|0, m_I\rangle$ and $|\pm 1, m_I\rangle$ states (Fig.1). A variety of similar multi-frequency resonances arising due to excited state LAC in a strong (51 mT) field have been studied in [12]. However, the resonances discovered in zero field in our work show some peculiarities: they appear as dips in the "normal" ODMR signal. Therefore, their nature must be similar to the "dark" resonances due to the coherent population trapping effect (CPT) in Λ-schemes [24]. These dips can be explained if we suppose that RF field does not drive forbidden transitions between levels with $\Delta m_S = \pm 2$, $\Delta m_I = 0$, but prevents these levels from interacting with MW drive field, as happens in CPT or EIT ([13,14]) schemes.

The multi-quantum spectra are also complicated by pure RF resonances described in [25]: an ultra-narrow (~7 KHz) line corresponding to the nuclear $|0,0\rangle \leftrightarrow |0,\pm 1\rangle$ transition was recorded in ODMR spectra in zero field under RF excitation at $f_{RF} = 4.95$ MHz. This line was located on an inclined substrate; both the line and the substrate give an additive contribution to the multi-quantum spectra shown on Fig.2,3.

## CONCLUSIONS

We report the detection of high-contrast magnetically independent two-quantum resonances in zero-field ODMR spectra of NV center in diamond, induced by applying an additional modulated RF field. These resonances can only be driven at $|B| < 0.5$ mT (in the case of our diamond sample), and therefore we can assert that they are due to the zero-field level anti-crossing. We attribute them to certain transitions in NV center's zero-field structure. We estimate the fractional short-term sensitivity of the timekeeping scheme based on these resonances in 0.01 mm$^3$ diamond sample in MW range as $\delta f/f \approx 1 \cdot 10^{-9}$ Hz$^{-1/2}$, which can be attractive for a wide class of metrological tasks.